\RequirePackage{fix-cm}
\documentclass[smallextended]{svjour3}       
\smartqed  
\usepackage{graphicx}
\begin{document}

\title{Spectroscopy of \boldmath $\eta'$-nucleus bound states at GSI and FAIR
}
\subtitle{--- very preliminary results and future prospects ---}

\titlerunning{Spectroscopy of $\eta'$-nucleus bound states at GSI and FAIR}        

\author{
H.~Fujioka$^{a}$ \and
Y.~Ayyad$^{b}$ \and
J.~Benlliure$^{c}$ \and
K.-T.~Brinkmann$^{d}$ \and
S.~Friedrich$^{d}$ \and
H.~Geissel$^{d,e}$ \and
J.~Gellanki$^{f}$ \and
C.~Guo$^{g}$ \and
E.~Gutz$^{d}$ \and
E.~Haettner$^{e}$ \and
M.N.~Harakeh$^{f}$ \and
R.S.~Hayano$^{h}$ \and
Y.~Higashi$^{i}$ \and
S.~Hirenzaki$^{i}$ \and
C.~Hornung$^{d}$ \and
Y.~Igarashi$^{j}$ \and
N.~Ikeno$^{k,l}$ \and
K.~Itahashi$^{m}$ \and
M.~Iwasaki$^{m}$ \and
D.~Jido$^{n}$ \and
N.~Kalantar-Nayestanaki$^{f}$ \and
R.~Kanungo$^{o}$ \and
R.~Knoebel$^{d,e}$ \and
N.~Kurz$^{e}$ \and
V.~Metag$^{d}$ \and
I.~Mukha$^{e}$ \and
T.~Nagae$^{a}$ \and
H.~Nagahiro$^{i}$ \and
M.~Nanova$^{d}$ \and
T.~Nishi$^{h}$ \and
H.J.~Ong$^{b}$ \and
S.~Pietri$^{e}$ \and
A.~Prochazka$^{e}$ \and
C.~Rappold$^{e}$ \and
M.P.~Reiter$^{e}$ \and
J.L.~Rodr\'{i}guez-S\'{a}nchez$^{c}$ \and
C.~Scheidenberger$^{d,e}$ \and
H.~Simon$^{e}$ \and
B.~Sitar$^{p}$ \and
P.~Strmen$^{p}$ \and
B.~Sun$^{g}$ \and
K.~Suzuki$^{q}$ \and
I.~Szarka$^{p}$ \and
M.~Takechi$^{r}$ \and
Y.K.~Tanaka$^{h}$ \and
I.~Tanihata$^{b,g}$ \and
S.~Terashima$^{g}$ \and
Y.N.~Watanabe$^{h}$ \and
H.~Weick$^{e}$ \and
E.~Widmann$^{q}$ \and
J.S.~Winfield$^{e}$ \and
X.~Xu$^{e}$ \and
H.~Yamakami$^{a}$ \and
J.~Zhao$^{g}$\\for the Super-FRS Collaboration
}

\authorrunning{H. Fujioka et al.} 

\institute{
H.~Fujioka \at
Kyoto University, Kitashirakawa-Oiwakecho, Sakyo-ku, 606-8502 Kyoto, Japan\\
\email{fujioka@scphys.kyoto-u.ac.jp} \and
$^{a}$Kyoto University, Kitashirakawa-Oiwakecho, Sakyo-ku, 606-8502 Kyoto, Japan \\
$^{b}$RCNP, Osaka University, 10-1 Mihogaoka, Ibaraki, 567-0047 Osaka, Japan \\
$^{c}$Universidade de Santiago de Compostela, 15782 Santiago de Compostela, Spain \\
$^{d}$Universit\"{a}t Giessen, Heinrich-Buff-Ring 16, 35392 Giessen, Germany \\
$^{e}$GSI, Planckstrasse 1, 64291 Darmstadt, Germany \\
$^{f}$KVI-CART, University of Groningen, Zernikelaan 25, 9747 AA Groningen, the Netherlands \\
$^{g}$Beihang University, Xueyuan Road 37, Haidian District, 100191 Beijing, China \\
$^{h}$The University of Tokyo, 7-3-1 Hongo, Bunkyo, 113-0033 Tokyo, Japan \\
$^{i}$Nara Women's University, Kita-Uoya Nishi-Machi, 630-8506 Nara, Japan \\
$^{j}$KEK, 1-1 Oho, Tsukuba, 305-0801 Ibaraki, Japan \\
$^{k}$Tohoku University, 6-3 Aoba, Aramaki, Aoba, Sendai, 980-8578 Miyagi, Japan\\
$^{l}$YITP, Kyoto University, Kitashirakawa-Oiwakecho, Sakyo-ku, 606-8502 Kyoto, Japan \\
$^{m}$Nishina Center, RIKEN, 2-1 Hirosawa, Wako, 351-0198 Saitama, Japan \\
$^{n}$Tokyo Metropolitan University, 1-1 Minami-Osawa, Hachioji, 192-0397 Tokyo, Japan \\
$^{o}$Saint Mary's University, 923 Robie Street, Halifax, Nova Scotia B3H 3C3, Canada \\
$^{p}$Comenius University Bratislava, FMFI, Mlynska dolina F1, 842 48, Bratislava, Slovakia\\
$^{q}$Stefan-Meyer-Institut f\"{u}r subatomare Physik, Boltzmangasse 3, 1090 Vienna, Austria \\
$^{r}$Niigata University, 8050 Ikarashi 2-no-cho, Nishi-ku, 950-2181 Niigata, Japan
}

\date{Received: date / Accepted: date}

\maketitle

\begin{abstract}
The possible existence of $\eta'$-nucleus bound states has been put forward
through theoretical and experimental studies.
It is strongly related to the $\eta'$ mass at finite density, which is expected to be reduced
because of the interplay between the $U_A(1)$ anomaly and partial restoration of chiral symmetry.
The investigation of the $\mathrm{C}(p,d)$ reaction at GSI and FAIR, as well as
an overview of the experimental program at GSI and future plans at FAIR are discussed.
\end{abstract}

\section{Introduction}
It is theoretically argued that the heavy mass of the $\eta'$ meson,
which is around $958\,\mathrm{MeV}/c^2$,
originates from
the interplay between the $U_A(1)$ anomaly in Quantum Chromodynamics and
the spontaneous breaking of chiral symmetry~\cite{Jido12}.
Furthermore, when chiral symmetry is partially restored at finite density,
the mass is expected to be reduced.
For example,
$150\,\mathrm{MeV}$ mass reduction at the normal nuclear density is estimated
in the Nambu--Jona-Lasinio model calculation~\cite{Nagahiro06},
and a recent calculation using the linear sigma model predicts
 $80\,\mathrm{MeV}$ reduction~\cite{Sakai13}. 
In the  QMC (quark-meson coupling) model,
a rather small mass reduction of $37\,\mathrm{MeV}$ is predicted
for the case of an $\eta$-$\eta'$ mixing angle of $-20^\circ$~\cite{Bass06}.

The mass reduction at finite density is, in other words,
the outcome of an attractive interaction between an $\eta'$ meson and a nucleus.
If the attraction is strong enough, an $\eta'$-nucleus bound state may be formed.
For the experimental observation of such a bound state,
the decay width is one of the important ingredients.
The chiral unitary model including a Lagrangian term 
which couples the singlet meson to the baryons 
indicates that the real part of the optical potential 
is in general deeper than the imaginary part,
while the coupling constant is an unknown parameter~\cite{Nagahiro12}. 

From the experimental side, two kinds of measurements,
$\eta'$ photoproduction off nuclei~\cite{Metag_EXA} and $\eta'$ production in proton-proton collision~\cite{Moskal_EXA}, have been used to extract
the information on $\eta'$-nucleus interaction and $\eta'$-nucleon interaction,
respectively, by taking different approaches.

The CBELSA/TAPS experiment determined both the real part~\cite{Nanova13} and the imaginary part~\cite{Nanova12} of the $\eta'$-nucleus optical potential.
The imaginary part, which is half of the absorption width,
 was obtained to be $10 \pm 2.5\,\mathrm{MeV}$ through a measurement of the transparency ratios
for different nuclear targets~\cite{Nanova12}.
In addition, by comparing the excitation function and the momentum distribution of $\eta'$ mesons
for the photoproduction off carbon with corresponding theoretical calculations,
the real part was derived as $-(37\pm 10\mbox{(stat)} \pm 10\mbox{(syst)})\,\mathrm{MeV}$~\cite{Nanova13}. 
Through these studies, they concluded a possible existence of $\eta'$ mesic nuclei~\cite{Metag_EXA,Nanova13}.

The near-threshold behavior of the cross section of the $pp\to pp\eta'$ reaction supplies 
information on the $\eta'$-nucleon interaction.
The COSY-11 experiment determined the $\eta'$-proton scattering length in free space,
as $(0\pm 0.43)+i(0.37^{+0.40}_{-0.16})\,\mathrm{fm}$.
It indicates a shallow depth of the $\eta'$-nucleus optical potential if one assumes 
no energy and density dependence~\cite{Moskal_EXA,Czerwinski14}.

The investigation of $\eta'$ mesic nuclei will be 
an alternative way to extract direct information of the $\eta'$-nucleus interaction.
We plan a series of missing-mass spectroscopy experiments with the $(p,d)$ reaction~\cite{Itahashi12}
at GSI and FAIR.
As the first step, we performed an inclusive measurement at GSI.
Moreover, a semi-exclusive measurement, with the coincidence of decay particles from $\eta'$ mesic nuclei,
will be studied in the future at FAIR. 

\section{Inclusive measurement at GSI}
\begin{figure}[t]
\begin{center}
\resizebox{\columnwidth}{!}{%
\includegraphics{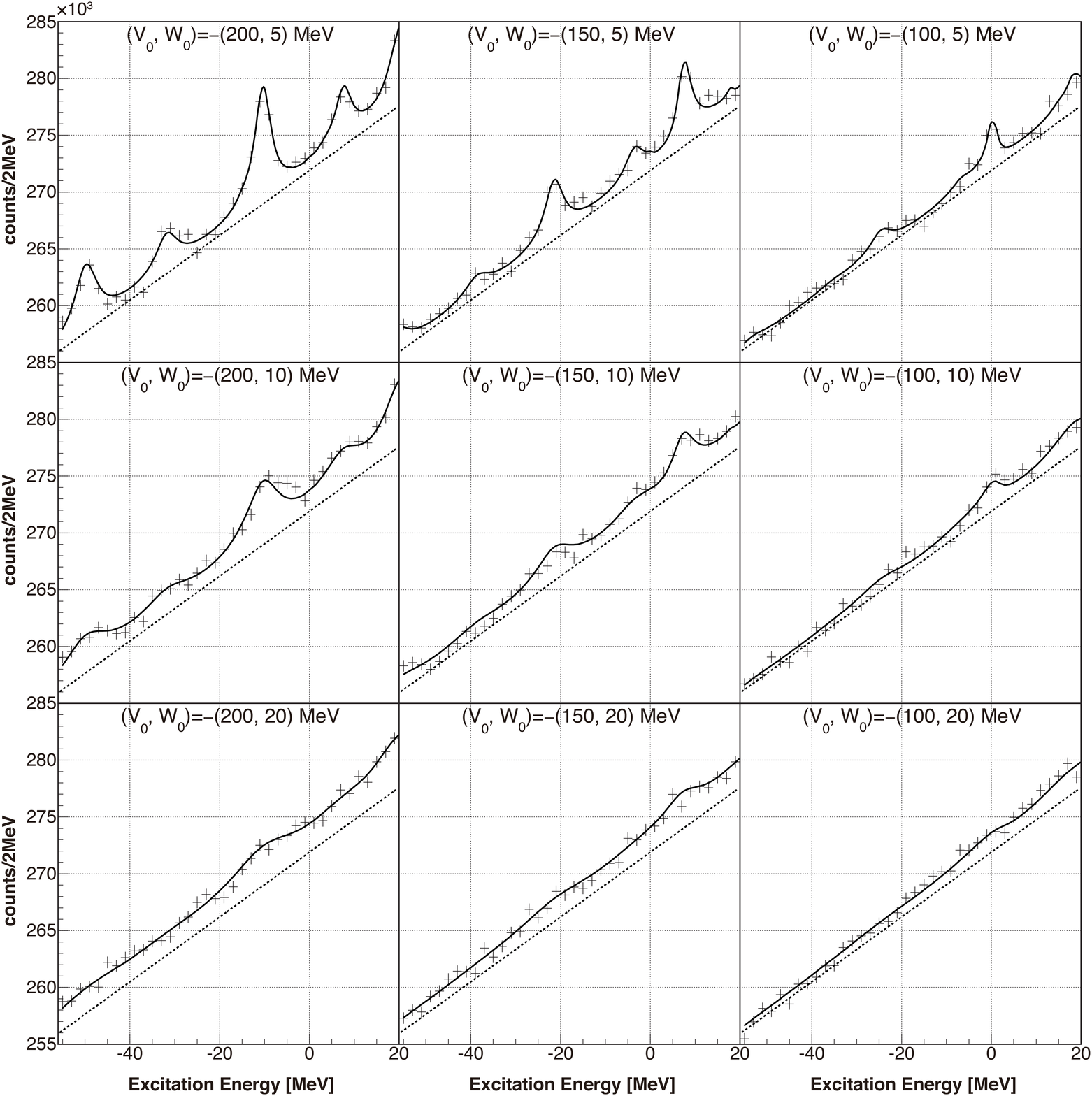} }
\caption{Simulated spectra with 4.5 days of data acquisition assuming different $\eta'$-nucleus optical potentials, parameterized as $V_0 + iW_0$. The dashed line corresponds to the background of quasi-free multi-pion production processes.}
\label{missingmass}       
\end{center}
\end{figure}

$\eta'$-mesic nuclei will be produced by impinging a $2.5\,\mathrm{GeV}$ proton beam 
onto a $4\,\mathrm{g/cm^2}$-thick carbon target.
The incident energy is slightly above the $\eta'$ production threshold in the $pn\to d\eta'$ reaction.
The missing-mass is obtained by analyzing the momentum of the ejectile deuterons with the Fragment Separator (FRS) used as a high-resolution spectrometer. 
The missing-mass resolution is estimated to be less than $2\,\mathrm{MeV}/c^2$ in $\sigma$,
which is dominated by the uncertainty of the reaction vertex inside the target.
It would be sufficiently smaller than the decay width of $\eta'$ mesic nuclei,
and this feature is one of the advantages of our experiment.

Figure~\ref{missingmass} shows a simulated spectrum corresponding to 4.5 days of data acquisition
for different $\eta'$-nucleus optical potentials parameterized as $V_0+iW_0$~\cite{Itahashi12}.
In the simulation, the cross section in the $(p,d)$ reaction is assumed 
to be the sum of a theoretical calculation~\cite{Nagahiro13,Nagahiro_EXA}, in which $\eta'$ production is involved,
and contributions of background processes of multi-pion production, 
evaluated from past measurements of proton-nucleon cross sections~\cite{ANKE} . 
The experimental sensitivity of finding peak structures above a huge background
will be large enough in case of a large $|V_0|$ and a small $|W_0|$.
While the signal-to-noise ratio will be of the order of 1/100 even in such a case, 
a high-statistics measurement with an intense primary beam
will compensate the poor signal-to-noise ratio. 
\begin{figure}[t]
\begin{center}
\resizebox{\columnwidth}{!}{%
\includegraphics{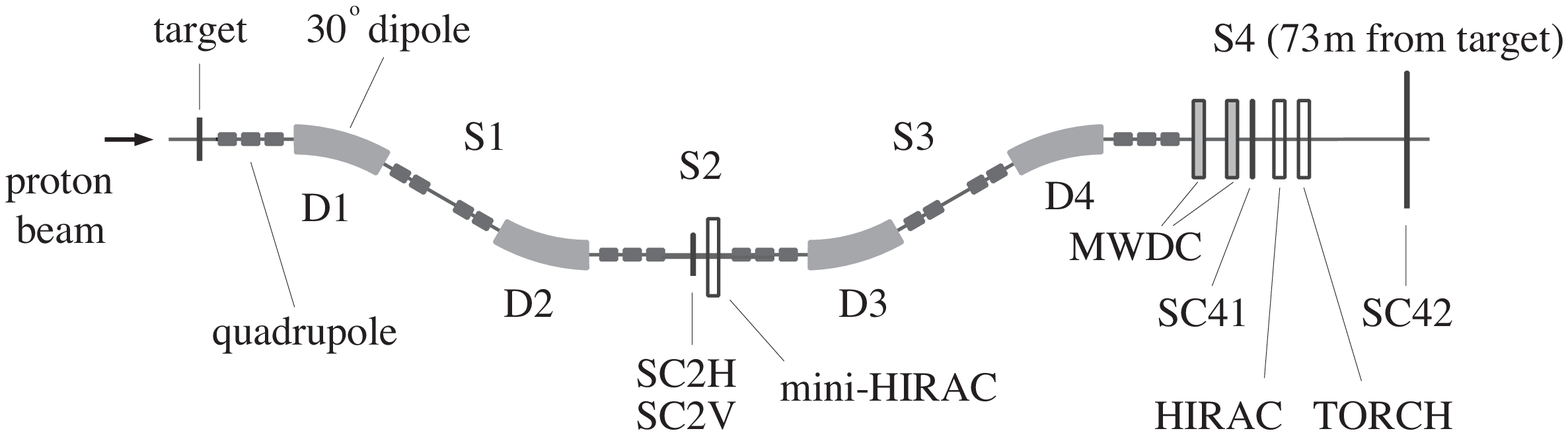} }
\caption{Schematic view of the experimental setup in August 2014.}
\label{frs_setup}       
\end{center}
\end{figure}

The first experiment (GSI S-437) was carried out at GSI in August 2014.
Figure~\ref{frs_setup} shows the experimental setup at FRS.
Two sets of multi-wire drift chambers (MWDC's) were installed close to a dispersive focal plane at S4,
and the momentum of ejectile deuterons could be analyzed by measuring their tracks with them.
For particle discrimination between deuterons and protons from inelastic scattering,
we measured the time-of-flight between S2 and S4
by plastic scintillators labelled with SC2H, SC2V, SC41, and SC42.
In addition, we prepared \v{C}erenkov detectors with high refractive-index ($n= 1.17\mbox{--}1.18$) silica aerogel radiators, which were developed at Chiba University~\cite{Tabata}, (HIRAC and mini-HIRAC) and a total-reflection \v{C}erenkov detector with an Acrylite radiator (TORCH).
It is worth emphasizing that we could make the main trigger for the $(p,d)$ reaction
with a tight coincidence of the signals of the scintillators in S2 and S4,
and without an anticoincidence of these \v{C}erenkov detectors.
This removes the influence of a possible position-dependence of the overkill rate of each \v{C}erenkov detector
on the missing-mass spectrum.

\begin{figure}[t]
\begin{center}
\resizebox{0.7\columnwidth}{!}{%
\includegraphics{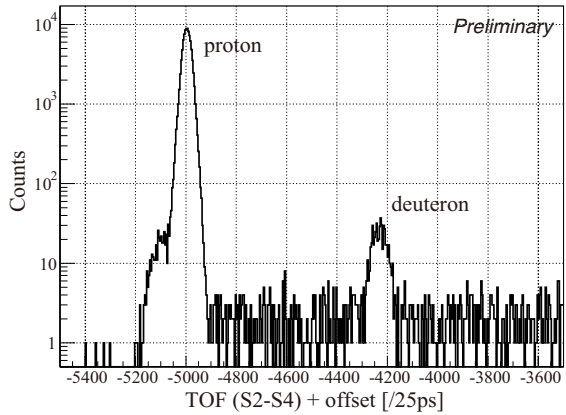} }
\caption{Time-of-flight distribution between S2 and S4.}
\label{tof2h41}       
\end{center}
\end{figure}

Figure~\ref{tof2h41} shows the TOF distribution between S2 and S4 obtained by an unbiased trigger.
The two peaks are $\sim 20\,\mathrm{ns}$ distant from each other, which is consistent with 
the TOF difference between protons and deuterons within the momentum acceptance.
The deuteron-to-proton ratio is found to be approximately 1/200.
Later on, we could achieve a deuteron-to-proton ratio around unity
by tuning the coincidence timing in order to remove the protons on the trigger level.
The remaining background, which is seen beneath the deuteron peak in Fig.~\ref{tof2h41}, is 
due to sequential protons with a short ($\sim 20\,\mathrm{ns}$) time interval. 
The waveforms of the scintillator signals will serve to distinguish one-pulse events from two-pulse events.
We have been working to improve the purity without losing the efficiency in particle identification.

We have taken data for the following reactions with changing the FRS magnetic field
 to cover a wide range of excitation energies between $-90\,\mathrm{MeV}$ and
 $+40\,\mathrm{MeV}$ relative to the $\eta'$ emission threshold.
\begin{itemize}
\item $^{12}\mathrm{C}(p,d)$ reaction at $2.5\,\mathrm{GeV}$ incident energy
\item $\mathrm{D}(p,d)$ reaction at $2.5\,\mathrm{GeV}$ incident energy (for background study)
\item $\mathrm{D}(p,d)p$ reaction at $1.6\,\mathrm{GeV}$ incident energy (for calibration)
\end{itemize}
The analysis of the MWDC's is in progress. 
For example, the overall missing-mass resolution for the $^{12}\mathrm{C}(p,d)$ reaction has been
estimated to be $\sigma \sim\,2\,\mathrm{MeV}$ (preliminary) from the missing-mass resolution
of the proton peak in the calibration process.

\section{Semi-exclusive measurement at FAIR}

While we will proceed with an inclusive measurement with higher statistics in a future experiment,
another way is to make a coincidence of the forward going deuteron
with decay particles from the decay of $\eta'$ mesic nuclei (semi-exclusive measurement).
We aim to tag a proton from the two-nucleon absorption process ($\eta'NN\to NN$),
which is one of the major decay modes of $\eta'$ mesic nuclei~\cite{Nagahiro13}.
This proton has a kinetic energy around $300\mbox{--}600\,\mathrm{MeV}$.
Such a fast proton will be hardly emitted from the main background of multi-pion production processes,
even if rescattering takes place,
and the coincidence is expected to improve the signal-to-noise ratio significantly.
More quantitative evaluation of the proton distribution from the background processes as well as 
the decay of $\eta'$ mesic nuclei, by an intra-nuclear cascade simulation using a microscopic transport model JAM~\cite{JAM}, is under way.

We plan to develop a proton counter to be installed surrounding the target, by which protons can be distinguished
from charged pions.
This experimental setup will be possible at Super-FRS, which is to be built in the FAIR complex.

\section{Summary}
We are investigating $\eta'$ mesic nuclei by means of missing-mass spectroscopy with the $(p,d)$ reaction.
An inclusive measurement of the $(p,d)$ reaction on ${}^{12}\mathrm{C}$ at FRS/GSI was carried out
in August 2014. The expected resolution will be around $2\,\mathrm{MeV}/c^2$, which is much smaller than the width of $\eta'$ mesic nuclei. The analysis is under way.

Furthermore, we plan to perform a semi-exclusive measurement at Super-FRS/FAIR,
with tagging decay protons from the two-nucleon absorption process in $\eta'$ mesic nuclei.
The signal-to-noise ratio, which is of the order of 1/100 at most in the inclusive spectrum, will be
improved significantly. A detailed Monte Carlo simulation with a microscopic transport model JAM is in progress.

\begin{acknowledgements}
The experiment was performed in the framework of the Super-FRS collaboration for FAIR.
This work is partly supported by a Grant-in-Aid for Scientific Research on Innovative Areas (No. 24105705)
from the Ministry of Education, Culture, Sports, Science and Technology(MEXT), Japan,
and a Grant-in-Aid for Young Scientists (A) (No. 25707018)
from Japan Society for the Promotion of Science (JSPS).
\end{acknowledgements}



\end{document}